\documentclass{article}

\usepackage{PRIMEarxiv}

\usepackage[utf8]{inputenc} 
\usepackage[T1]{fontenc}    
\usepackage{hyperref}       
\usepackage{url}            
\usepackage{booktabs}       
\usepackage{amsfonts}       
\usepackage{nicefrac}       
\usepackage{microtype}      
\usepackage{lipsum}
\usepackage{fancyhdr}       
\usepackage{graphicx}       
\graphicspath{{media/}}     
\usepackage{float}
\usepackage{amsmath}

\usepackage{placeins} 

\title{Unveiling the Bright Spot with Mie Scattering}

\author{
  Vinícius dos Passos de Souza, Antonio Alvaro Ranha Neves \\
  Federal University of the ABC (UFABC) \\
  Santo André\\
  passsos.souza@ufabc.edu.br}

\begin{document}
\maketitle

\begin{abstract}
This work applies Mie scattering theory to provide a new perspective on the propagation of light through a spherical obstacle, offering a novel explanation for the formation of the Poisson spot (also known as the Arago or Fresnel spot). We demonstrate that the diffraction patterns generated by a sphere and by a circular disk can be understood as complementary outcomes of the same underlying scattering process. Our analysis highlights the constructive interference responsible for the bright central spot, and extends the classical wave optics framework by connecting it directly with the scattering coefficients of spherical harmonics. This approach not only deepens the theoretical understanding of diffraction phenomena, but also provides a practical framework that may be applied in modern optical experiments and photonic device design.
\end{abstract}


\section{Introduction}
The diffraction of light has long been a cornerstone of classical wave optics, revealing the wave nature of light through interference and scattering phenomena. One of the most striking manifestations of diffraction is the Poisson spot, also referred to as the Arago\cite{SpotArago,BookHist} or Fresnel spot\cite{FresnelSpot}, a bright point that appears at the center of the shadow cast by a circular obstacle. First predicted by Poisson as a direct challenge to Fresnel Theory of wave light \cite{fresnel1816} and experimentally confirmed by Dominique Arago \cite{arago1819} in the early 19th century, this phenomenon challenged the particle theory of light and strongly supported the wave theory.

To remain neutral regarding the contributions of these pioneers, we will adopt the term ``The Bright Spot'', as suggested in a problem from the 2024 International Physicists' Tournament \cite{IPT}.

\begin{figure}[h]
    \centering
    \includegraphics[width=\linewidth]{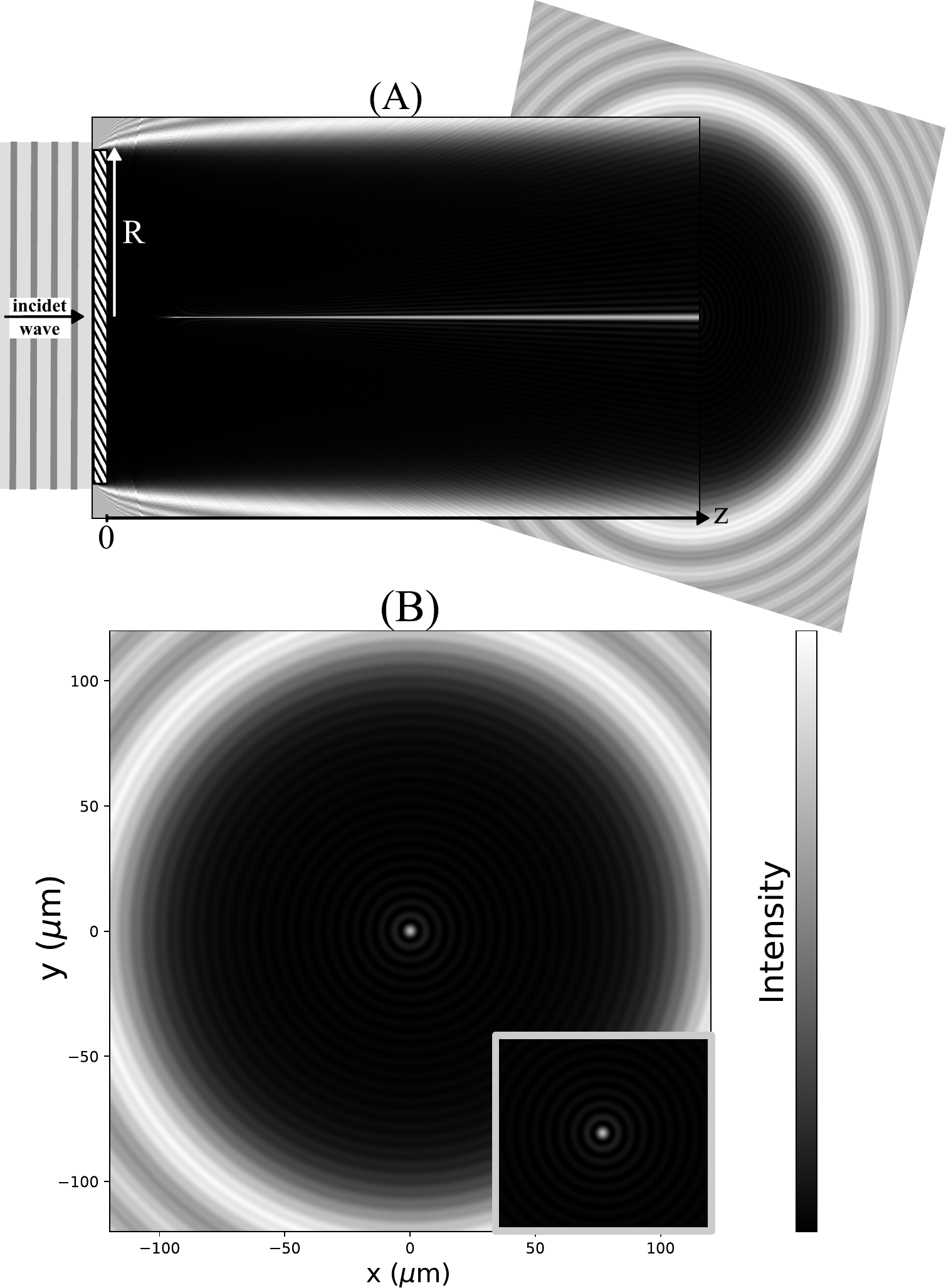}
    \caption{Schematic of light propagation around a circular disk, with $R =$15~{\textmu}m (A) and the resulting diffraction pattern featuring the Bright Spot at a distance $z =$1800$\cdot$ R {\textmu}m, with zoom in the center region to show bright spot (B). Images generated using Fourier Optics.}
    \label{fig:introduc}
\end{figure}

Beyond its historical significance, the Bright Spot phenomenon remains relevant across numerous scientific and engineering fields. In quantum physics, it is crucial for demonstrating interference in molecular beams and probing van der Waals and Casimir--Polder interactions \cite{Juffmann2012,PhysRevA.94.023621,PhysRevA.96.013626,D2CP03349F}. In optical metrology, the spot's sensitivity is leveraged for alignment systems in laser fusion experiments, free-electron lasers, and particle colliders \cite{Saruta_2007,Sakauchi2009,He_2022,Griffith1990,LEBEDEV2002163}. Furthermore, in astronomy, the Bright Spot is central to the design of coronagraphs for exoplanet detection \cite{Vanderbei_2007,Shiri_2013}. Given the prevalence of circular and spherical obstacles in these applications, Mie scattering theory offers a precise and comprehensive framework for modeling the underlying diffraction phenomena. Furthermore, lithography and nanofabrication techniques use diffraction from opaque disks to create microtube arrays and subwavelength patterns \cite{Jung_2012,Tian_2011}. Similarly, X--ray diffraction employs analogous phenomena involving constructive interference in the shadow of circularly symmetric obstacles \cite{Evans_2010}. Given the prevalence of circular and spherical obstacles in these applications, Mie scattering \cite{gouesbet2023generalized} theory offers a precise and comprehensive framework for modeling the underlying diffraction phenomena.

An analytical solution for the Bright Spot produced by a circular disk, derived from the Rayleigh--Sommerfeld formulation, is well-established for radius values inside of the disk \cite{Analy}:

\begin{equation}
    I(\rho) = \frac{A^2 z^2}{z^2+R^2}J_0^2 \left( r \frac{2R\pi}{\lambda z}\right),
    \label{eq:analytcirc}
\end{equation}
\FloatBarrier
where $\lambda$ is the wavelength, $z$ is the distance from the disk to the observation plane, $R$ is the disk's radius, and $r$ is the radial distance from the center of the disk on the xy plane, as illustrated in Fig. \ref{fig:introduc}.

While the diffraction from a 2D disk is well understood, the case of a 3D sphere is less explored in this context. Although the Mie Scattering, has been used to study phenomena like photonic nanojets \cite{Antonio}, it has not been applied to analyze the Bright Spot from opaque spheres or even used to discuss the appearing of this Bright Spot

In this work, we use traditional diffraction theory for the disk and extend the analysis to a sphere using Mie theory. This dual approach provides a new explanation for the Bright Spot's formation and offers a method to distinguish between 2D (disk) and 3D (sphere) objects based on their diffraction patterns.

\section{Theory}

Mie scattering, first introduced by Gustav Mie \cite{Mie}, provides an exact analytical solution to Maxwell's equations for the scattering of electromagnetic radiation by a spherical particle. Starting from the scalar wave equation in spherical coordinates, the general solution can be written as:

\begin{equation}
    U(r) = \sum_{n,m} \left ( A_{n,m}z_n(kr) + B_{n,m}z_n(kr)\right) Y_{n,m}(\theta,\phi),
    \label{eq:wave}
\end{equation}

where $A_{n,m}$ and $B_{n,m}$ are constants determined by boundary conditions, $z_n$ represents spherical Bessel or Hankel functions, and $Y_{n,m}$ are the spherical harmonics. Assuming a time--harmonic dependence of $\exp(-\text{i}\omega t)$, which is suppressed hereafter, the electromagnetic fields can be derived by applying the angular momentum operator, $\vec{L}$. This defines Transverse Electric (TE) and Transverse Magnetic (TM) modes, allowing the fields to be expressed as \cite{Antonio}:

\small
\begin{align}
    \textbf{E} &= E_0 \sum_{n,m}\left( \frac{\mathrm{i}}{k}A_{n,m}\nabla\times z_n(kr) \textbf{X}_{n,m}(\theta,\phi) + B_{n,m}z_n(kr)\textbf{X}_{n,m}(\theta,\phi) \right), \\
    \textbf{H} &= H_0 \sum_{n,m} \left( A_{n,m} z_n(kr) \textbf{X}_{n,m}(\theta,\phi) - \frac{\mathrm{i}}{k} B_{n,m}\nabla\times z_n(kr)\textbf{X}_{n,m}(\theta,\phi) \right),
\label{eq:ET&MTsol1}
\end{align}
\normalsize

where \(\textbf{X}_{n,m}\) is the harmonic spherical vector. And to apply the boundary conditions of the problem we will need to describe the incident, scattered and inside fields. Which can be made by using spherical Bessel functions for the incident and inside fields, giving their place in space, and the scattered field using the Hankel function. In each of them we will have new constants, where the incident field once defined can have it's constant calculated. Thus leading us to write the incident fields:
\small
\begin{equation}
    \textbf{E}_{\text{inc}}= E_0 \sum_{n,m}\left( \frac{\mathrm{i}}{k}G_{n,m}^{\text{TM}}\nabla \times j_n(k r)\textbf{X}_{n,m}(\theta,\phi) + G_{n,m}^{\text{TE}}j_n(k r)\textbf{X}_{n,m}(\theta,\phi) \right),
    \label{E_inc0}
\end{equation}
\begin{equation}
    \textbf{H}_{\textit{inc}} = \frac{E_0}{Z}\sum_{n,m} \left(G_{n,m}^{\text{TM}}j_n(k r)\textbf{X}_{n,m}(\theta,\phi) -\frac{\mathrm{i}}{k} G_{n,m}^{\text{TE}}\nabla \times j_n(k r)\textbf{X}_{n,m}(\theta,\phi) \right).
    \label{B_inc0}
\end{equation}
\normalsize
where $\textbf{X}_{n,m}$ is the vector spherical harmonic. While the scattered field is described using spherical Hankel functions, given by:
\small
\begin{equation}
    \textbf{E}_{\text{sca}} = E_0 \sum_{n,m}\left( \frac{\mathrm{i}}{k}a_{n,m}\nabla \times h_n^{(1)}(k r)\textbf{X}_{n,m}(\theta,\phi) + b_{n,m}h_n^{(1)}(k r)\textbf{X}_{n,m}(\theta,\phi) \right),
    \label{E_scat0}
\end{equation}
\begin{equation}
    \textbf{H}_{\text{sca}} = \frac{E_0}{Z_1}\sum_{n,m} \left(a_{n,m}h_n^{(1)}(kr)\textbf{X}_{n,m}(\theta,\phi) -\frac{\mathrm{i}}{k} b_{n,m}\nabla \times h_n^{(1)}(kr)\textbf{X}_{n,m}(\theta,\phi) \right).
\end{equation}
\normalsize
The internal fields have a similar form but with different coefficients ($c_{n,m}$ and $d_{n,m}$). By applying the continuity boundary conditions at the sphere's surface, we can solve for the scattering coefficients, $a_{nm}$ and $b_{nm}$ \cite{Ranha_Neves}. For a dielectric sphere, these coefficients depend on the size parameter $x = kR$ (where $k$ is the wavenumber and $R$ is the sphere radius) and the relative refractive index $M$, defined as the ratio of the inside refractive index over the outside.

\begin{equation}
    a_{n,m} = \frac{M\Psi_n(Mx)\Psi_n^{\prime}(x)-\Psi_n^{\prime}(Mx)\Psi_n(x)}{\Psi_n^{\prime}(Mx)\zeta_n(x)-M\Psi_n(Mx)\zeta_n^{\prime}(x)}G_{n,m}^{\text{TM}}
    \label{eq:MieCoefA}
\end{equation}
\begin{equation}
    b_{n,m} = \frac{M\Psi_n^{\prime}(Mx)\Psi_n(x)-\Psi_n(Mx)\Psi_n^{\prime}(x)}{\Psi_n(Mx)\zeta_n^{\prime}(x)-M\Psi_n^{\prime}(Mx)\zeta_n(x)}G_{n,m}^{\text{TE}},
    \label{eq:MieCoef}
\end{equation}

In this work, we model the opaque obstacle as a perfectly conducting sphere. This simplifies the boundary conditions, as described in \cite{PerfCond}, leading to the following Mie coefficients:

\begin{equation}
    a_{n,m} = -\frac{\Psi_n'(x)}{\zeta_n'(x)}G_{n,m}^{\text{TM}}=a_n G_{n,m}^{\text{TM}}
\end{equation} 

\begin{equation}
    b_{n,m} = - \frac{\Psi_n(x)}{\zeta_n(x)}G_{n,m}^{\text{TE}}=b_nG_{n,m}^{\text{TE}}.
\end{equation}

where the prime ($'$) denotes the derivative with respect to the argument, and the Riccati--Bessel functions are $\Psi_n(\rho) = \rho j_n(\rho)$ and $\zeta_n(\rho) = \rho h_n^{(1)}(\rho)$. Notably, these expressions do not depend on the material properties ($M$) of the sphere. With these coefficients, or any other Mie coefficients for real instead of a perfect conductor, the total electric field, and thus the light intensity, squared modulus of the total electric field, can be calculated at any point in space.

\section{Methods}

A Python script was developed to calculate the light intensity pattern based on the derived equations. The script takes the sphere radius ($R=$100~{\textmu}m), illuminated by light with $\lambda=$0.632~{\textmu}m, the number of terms in the Mie series, and the spatial coordinates as inputs. For all simulations, the infinite Mie series was truncated after 2000 terms to minimize truncation error \cite{TruncationError}. The incident wave was defined as plane waves polarized on the x direction, defining the values of the Mie constants, presented in the Supplementary Material.

For all of the $z$ distances presented the zero is on the surface of the object, for a disk is at zero, but for a sphere is at the $z = R$.

The source code used for these calculations is publicly available on \href{https://github.com/PassosSouza/MieBrightSpot}{GitHub}.

\section{Result and discussion}

\begin{figure}[htb]
    \centering
    \includegraphics[width=\linewidth]{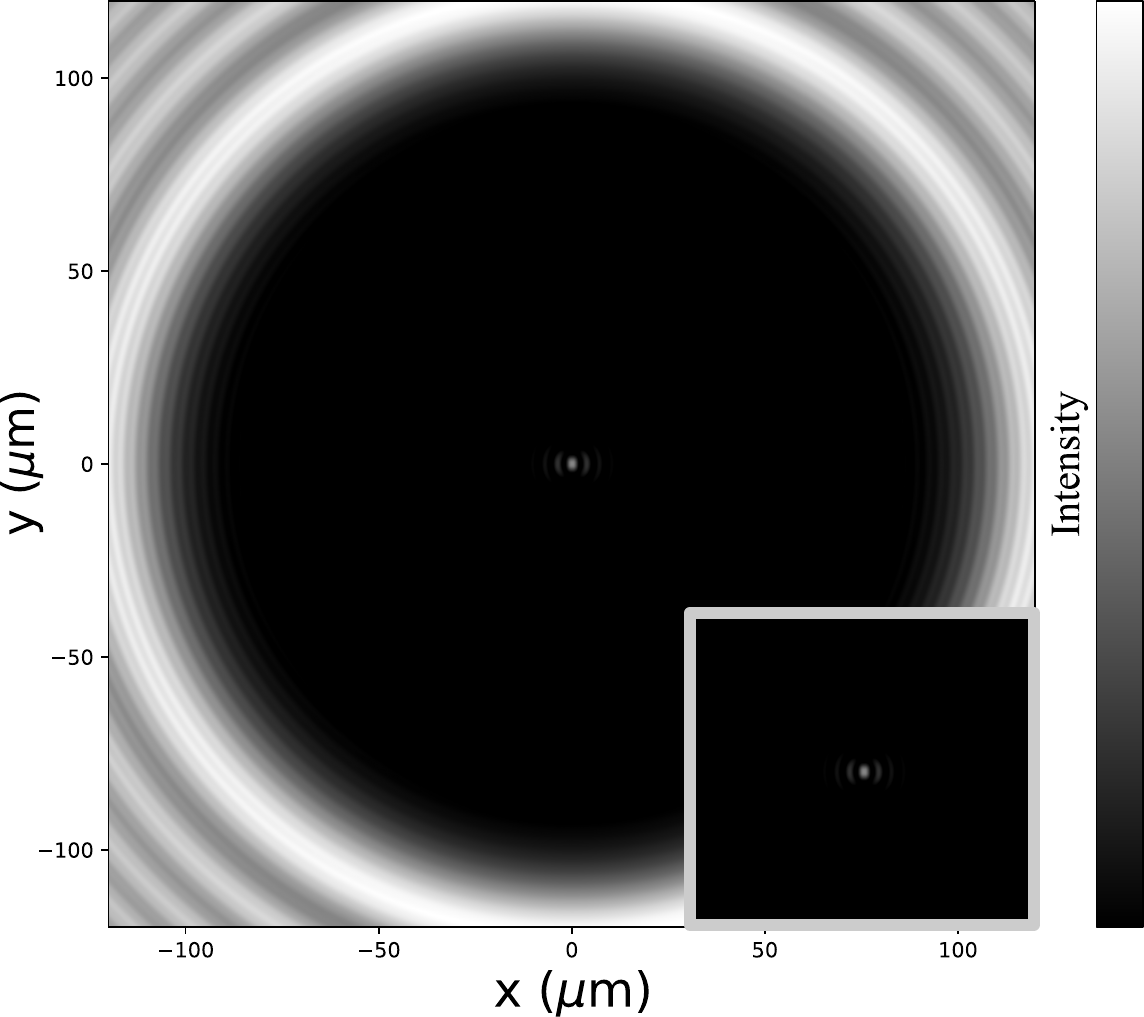}
    \caption{Calculated diffraction pattern from a perfectly conducting sphere using Mie theory, showing a central Bright Spot. Parameters, as indicated in the Methods section, and observation plane at $z =$ 1000 {\textmu}m from the sphere's edge.}
    \label{fig:Spot}
\end{figure}

Using the Mie scattering model for a perfectly conducting sphere, we calculated the diffraction pattern and observed a distinct bright spot at the center of the shadow, as shown in Fig. \ref{fig:Spot}. This result confirms, as expected, that Mie theory successfully describes this classic diffraction phenomenon for 3D objects, with the Bright Spot appearing at the center of the shadow.

Crucially, the pattern reveals a subtle but significant difference in the central spot's geometry, appearing slightly deformed along the x-axis, which is the axis of the incident field's polarization. This deformation is a direct physical manifestation of the vector nature of the light-scattering interaction. Unlike scalar diffraction theories, the rigorous Mie solution inherently accounts for the complex coupling of Transverse Electric (TE) and Transverse Magnetic (TM) modes, defined by the Mie coefficients and the incident field constants ($G_{n,m}^{\text{TE}}$ and $G_{n,m}^{\text{TM}}$). This capability proves that the diffraction process involves a vector scattering mechanism that cannot be adequately described by scalar wave optics, establishing the theoretical superiority of the Mie solution for analyzing this phenomenon.

\FloatBarrier

\begin{figure}[h]
    \centering
    \includegraphics[width=\linewidth]{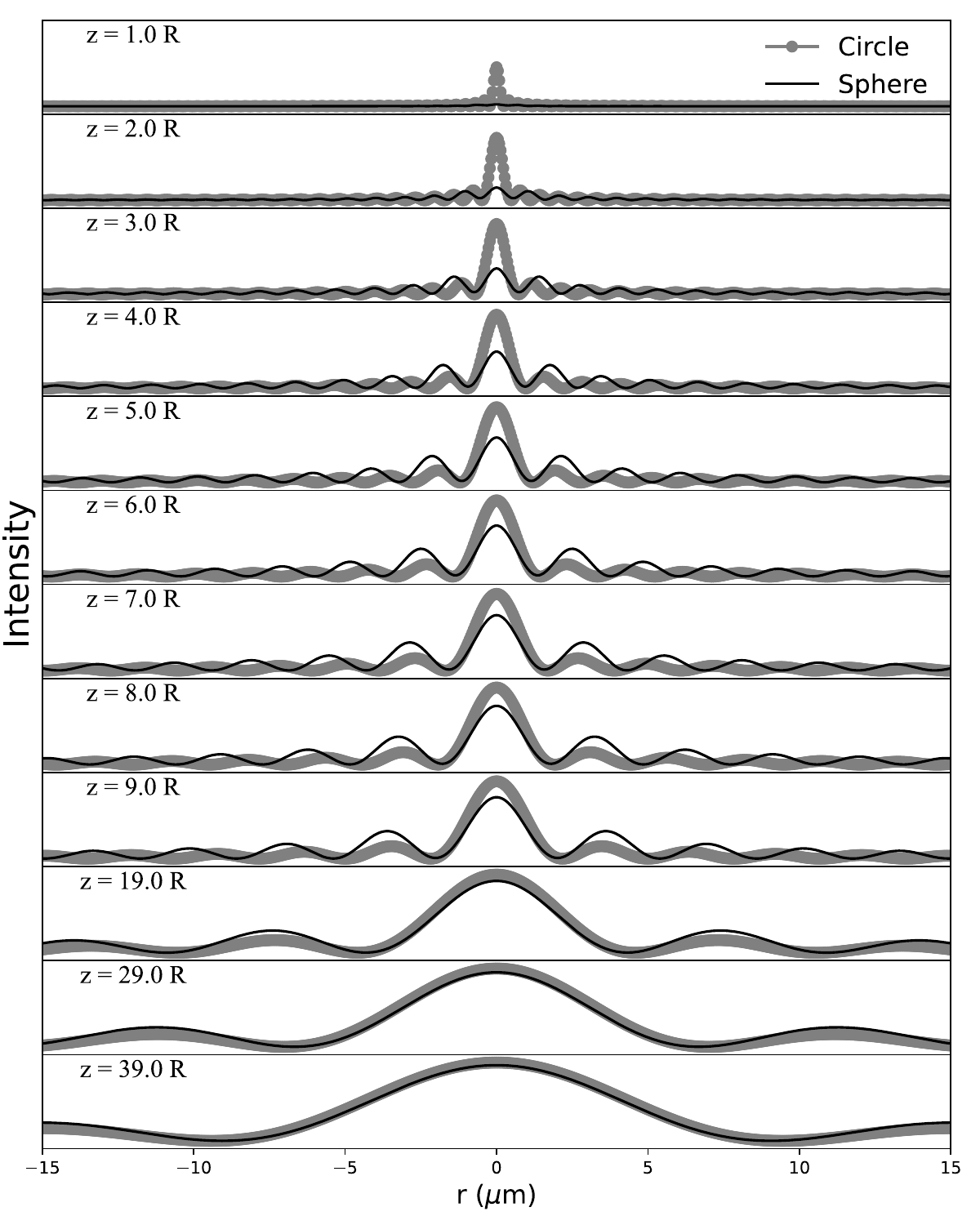}
    \caption{Comparison of intensity profiles for a circular disk (analytical solution) and a sphere (Mie theory). Both objects have same radius $R=$100 {\textmu}m, as indicated in the Methods section. Profiles are shown at multiple distances ($z$) from the object's edge.}
    \label{fig:Compara1}
\end{figure}

But now appears a question, is the profile seen on the disk the same as on the sphere?
To answer it, we can use the models for both sphere (Mie theory) and circular disk (Eq.~\ref{eq:analytcirc})), we can compare their diffraction patterns. Figure~\ref{fig:Compara1}  shows the intensity profiles at different distances ($z$) from the obstacle. For small values of $z$, the patterns are noticeably different on the intensity measured, but similarly on the peak positions seen. As $z$ increases, the patterns become more similar, indicating that far from the obstacle, it becomes increasingly difficult to distinguish between the sphere and the disk based on the overall diffraction pattern alone.
To find a reliable method for distinguishing the two objects, we analyzed the on-axis intensity as a function of distance $z$. Because of that, we only need to know the intensity at the center point, where \(\theta=0\), so instead of using the general equation, we simplified it for this specific case, with all calculations shown in the Supplementary material. Thus allowing us to write the scattered field in a simpler equation.


Revealing that the intensity of the light depends on the \(h_n^{(1)}(kz)\) and Mie Coeffcients for a perfectly conductive sphere. For each spherical harmonic order, n, the maximum intensity contribution of the spherical Hankel function is located at an axial distance z where the argument \(kz\) is near n. Looking at the \(j_n(kz)\), which are the real part of the \(h_n^{(1)}(kz)\), follows the same principle, however its maximum value decreases as the order n increases. Consequently, at any given axial distance z, the total intensity is predominantly shaped by the low-order terms whose n values are closest to kz. This intricate, order-dependent behavior fundamentally contrasts with the center intensity calculated for a circular disk, Eq. \ref{eq:analytcirc}, which relies simply on the inverse square of the radius, clearly indicating the source of the expected differences in the diffraction patterns.

\FloatBarrier

\begin{figure}[htb]
    \centering
    \includegraphics[width=\linewidth]{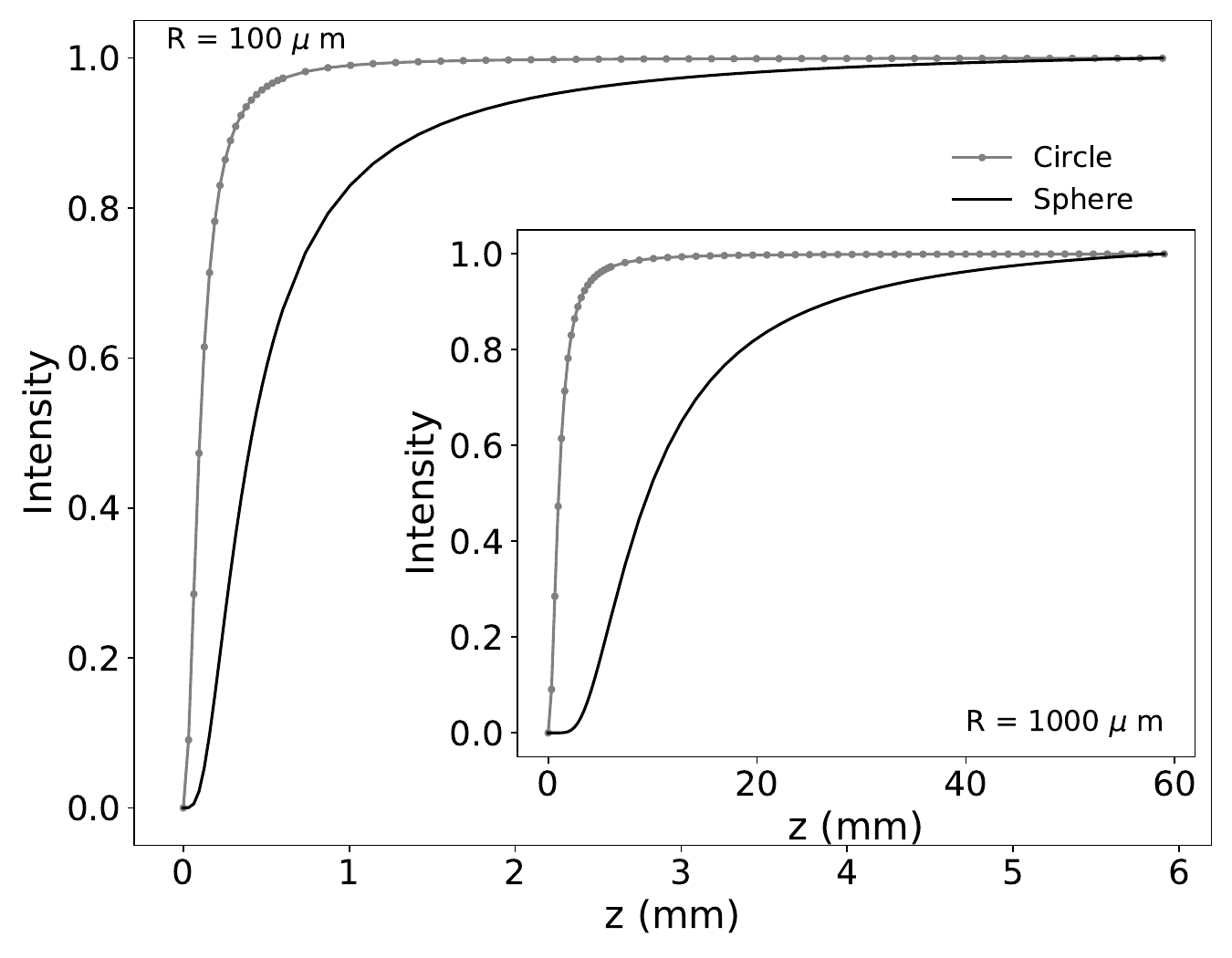}
    \caption{Calculations of on--axis intensity of the Bright Spot as a function of distance ($z$) from the object's edge for a disk and a sphere. The differing profiles offer a clear method for distinguishing the two geometries.}
    \label{fig:Compara2}
\end{figure}

We compared the on-axis intensity profile of a sphere with that of a disk shown in Fig. \ref{fig:Compara2}. The profiles for the sphere and the disk are clearly different, particularly at shorter distances from the object, where a less inclined increase on intensity can be seen, generating a really different curve from the disk and at bigger radius a plateau can be seen near zero initially.

This difference  of the on axis intensity likely arises because a sphere can be thought of as a stack of disks with varying radius. At closer distances, the scattering contributions from these multiple layers become more pronounced, leading to reduced in-phase interference compared to a single disk.

The difference between the disk and the sphere becomes increasingly pronounced as the object’s radius grows. This distinct behavior of the on-axis intensity provides a measurable parameter that can be used experimentally to distinguish between spherical and circular obstacles. Moreover, the variation in intensity between characteristic points becomes more evident for larger radius, further highlighting the contrast between the two cases. In the Supplementary Material, we present a detailed comparison over multiple distances, including the points where the intensity first rises from zero and where it reaches its maximum value. While these points do not directly yield the exact radius of the sphere, they clearly demonstrate the significant differences in behavior between spheres and disks.

\FloatBarrier

\section{Conclusion}

We successfully applied the established Mie scattering theory to rigorously describe the formation of the Bright Spot for a perfectly conducting sphere. This work presents a novel application of Mie theory in this context, establishing a rigorous, vector--based framework for analyzing diffraction by 3D spherical objects, in direct comparison to traditional scalar--based models for 2D circular disks.

Our analysis confirms the appearance of the Bright Spot within the framework of Mie theory and, more importantly, reveals a subtle polarization--dependent deformation of the central spot, a direct manifestation of the vector nature of light-scattering interactions that scalar theories fail to capture.

Furthermore, we demonstrated that the on--axis intensity profile as a function of distance provides a distinct signature for each geometry. Measuring this intensity is a clear and effective experimental method to distinguish between a circular disk and a sphere. This distinction is more pronounced at shorter distances and for larger radii, which can be leveraged for accurate object classification in various optical applications.

In summary, this approach provides a deeper theoretical understanding by integrating this classic diffraction phenomenon with the full vector solution of Mie scattering and offers a powerful diagnostic tool for object characterization in fields such as optical metrology and nanofabrication.

\section*{Acknowledgments}
Vinicius acknowledges the financial support from FAPESP (Grants 2019/22183-6 and 2024/16253-0).

\section*{Data Availability Statement}

The authors have no conflicts to disclose; all codes used are available on \href{https://github.com/PassosSouza/MieBrightSpot}{https://github.com/PassosSouza/MieBrightSpot}

\FloatBarrier  
\nocite{*}

\bibliographystyle{unsrt}  
\bibliography{references}  

\newpage

\section*{Appendice}
\subsection*{Fourier Optics}

For the Fourier optics simulations, we employed a standard method \cite{Livro}. Starting from a spatial grid that defines the optical field distribution \(u_0(x,y,0)\) at the initial plane, its angular spectrum is obtained through the two-dimensional Fourier transform:

\begin{equation}
    A(k_x,k_y) = \int_{-\infty}^{\infty}\int_{-\infty}^{\infty} 
    u_0(x,y,0)\, \exp{\left[-i(k_x x + k_y y)\right]} \, \mathrm{d}x\, \mathrm{d}y .
\end{equation}

This representation allows the calculation of the propagated field at a distance \(Z\) along the optical axis, given by

\begin{equation}
    F(x,y,Z) = \int_{-\infty}^{\infty}\int_{-\infty}^{\infty} 
    A(k_x,k_y)\, \exp{\left(\,iZ\sqrt{k^2 - k_x^2 - k_y^2}\right)} 
    \, \exp{\left[\,i(k_x x + k_y y)\right]} \, \mathrm{d}k_x\, \mathrm{d}k_y ,
\end{equation}

which corresponds to the inverse Fourier transform of the angular spectrum multiplied by a propagation phase factor.  

This procedure was implemented numerically in Python using the \texttt{SciPy} package for the Fourier and inverse Fourier transforms. The simulations were performed on grids of size \(N \times N\), with \(N = 3072\), and with a computational window chosen to be nine times larger than the beam radius, in order to minimize boundary effects.  

\begin{figure}[h]
\centering
\includegraphics[width=1\linewidth]{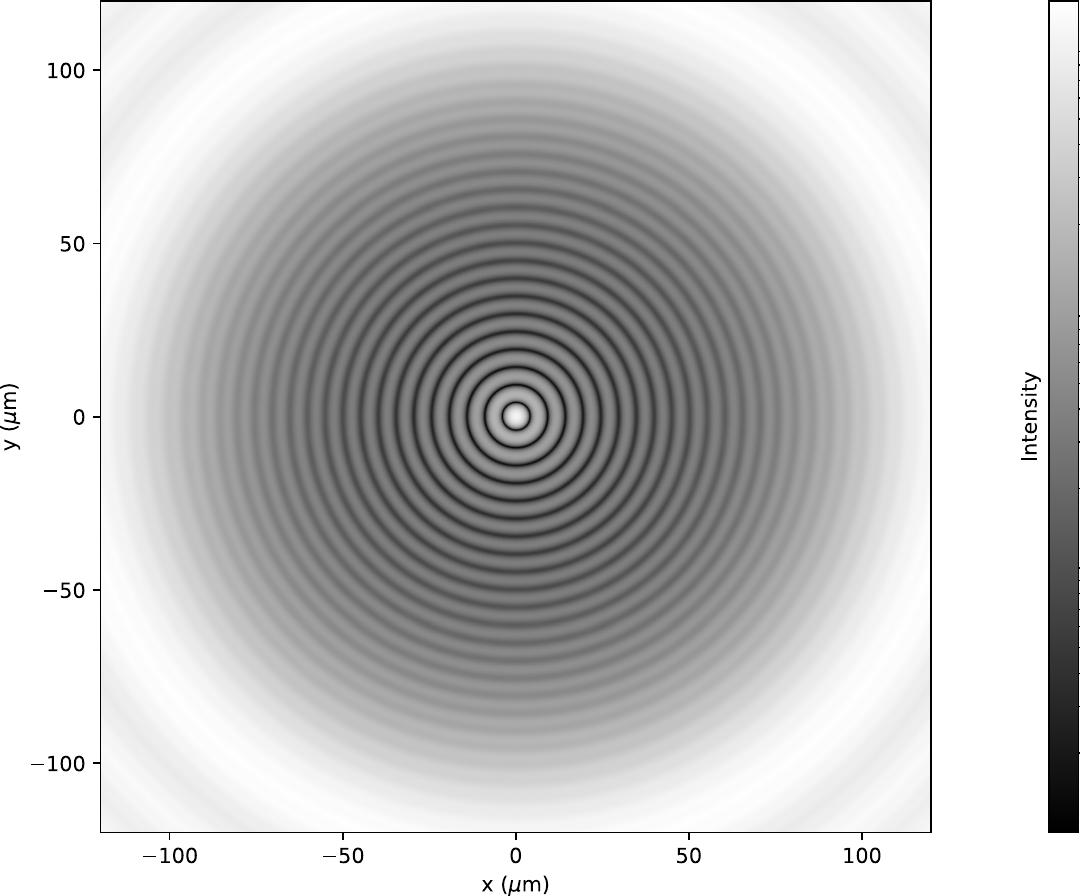}
\caption{The simulation of the pattern generated by a disk with a \(R = 100\) \textmu m, the same patter and in Figure 1 (B) in the text, but in log scale.}
\label{fig:IntroducLog}
\end{figure}

The image was made and to make it easier to see the multiple lines expected in here there is a version of the Figure 1 (B) in log scale, Figure \ref{fig:IntroducLog}.

The source code used for the simulations is openly available at:  
\href{https://github.com/PassosSouza/MieBrightSpot}{GitHub}

\newpage

\subsection{Mie Coefficients for an Incident Field}

We define the incident electric and magnetic fields in terms of vector spherical harmonics as
\begin{equation}
    \textbf{E}_{\text{inc}} = E_0 \sum_{n,m} \left( 
        \frac{\mathrm{i}}{k} G_{n,m}^{\text{TM}} \nabla \times \big[ j_n(kr)\, \textbf{X}_{n,m}(\theta,\phi) \big] 
        + G_{n,m}^{\text{TE}} j_n(kr)\, \textbf{X}_{n,m}(\theta,\phi) 
    \right),
    \label{eq:E_inc}
\end{equation}
\begin{equation}
    \textbf{H}_{\text{inc}} = \frac{E_0}{Z} \sum_{n,m} \left(
        G_{n,m}^{\text{TM}} j_n(kr)\, \textbf{X}_{n,m}(\theta,\phi) 
        - \frac{\mathrm{i}}{k} G_{n,m}^{\text{TE}} \nabla \times \big[ j_n(kr)\, \textbf{X}_{n,m}(\theta,\phi) \big] 
    \right),
    \label{eq:H_inc}
\end{equation}
where \(j_n(kr)\) are spherical Bessel functions of the first kind, and \(\textbf{X}_{n,m}(\theta,\phi)\) are vector spherical harmonics.  

The expansion coefficients can be obtained from \cite{Antonio}:
\begin{equation}
    G_{n,m}^{\text{TE}} = \frac{-k}{z_n(kr)\, E_0\sqrt{n(n+1)}}  
    \int Y_{n,m}^*(\theta,\phi)\, \textbf{r}\cdot \textbf{H}_{\text{inc}}\, \mathrm{d}\Omega ,
\end{equation}
\begin{equation}
    G_{n,m}^{\text{TM}} = \frac{k}{z_n(kr)\, H_0\sqrt{n(n+1)} } 
    \int Y_{n,m}^*(\theta,\phi)\, \textbf{r}\cdot \textbf{E}_{\text{inc}}\, \textrm{d}\Omega ,
\end{equation}
where \(z_n(kr)\) can represent spherical Bessel functions of the first kind (\(j_n\)), second kind (\(y_n\)), or spherical Hankel functions (\(h_n^{(1)}, h_n^{(2)}\)), depending on the problem symmetry.  
The integration measure is \(\mathrm{d}\Omega = \sin\theta~ \mathrm{d}\theta\, \mathrm{d}\phi\), and the unit vector is expressed as
\[
    \textbf{r} = \cos\phi \sin\theta \,\textbf{x} + \sin\phi \sin\theta \,\textbf{y} + \cos\theta \,\textbf{z}.
\]
The spherical harmonics are defined as
\begin{equation}
    Y_{n,m}^*(\theta,\phi) = (-1)^m 
    \sqrt{\frac{2n+1}{4\pi}\cdot \frac{(n-m)!}{(n+m)!}}
    \, P_n^m(\cos\theta)\, \exp{\left(-\textrm{i} m \phi\right)}.
\end{equation}

\subsection{Mie Coefficients for a Plane Wave}

For a plane wave propagating along the \(z\)-axis, polarized along \(\textbf{x}\), we have
\begin{equation}
    \textbf{E} = E_0\, e^{i k z}\, \textbf{x}, \qquad
    \textbf{H} = H_0\, e^{i k z}\, \textbf{y}.
\end{equation}
In this case, we select the spherical Bessel function of the first kind, \(j_n(kr)\). Substituting into the integral expression gives
\begin{equation}
\begin{aligned}
    G_{n,m}^{\text{TE}} &= (-1)^m\frac{k r \,}{j_n(kr)\,\sqrt{n(n+1)} } 
    \sqrt{\frac{2n+1}{4\pi}\cdot \frac{(n-m)!}{(n+m)!}} \\
    &\times \int_0^\pi P_n^m(\cos\theta)\, \exp{\left(\mathrm{i} k r \cos\theta\right)} \sin^2\theta\, \mathrm{d}\theta 
    \int_0^{2\pi} \exp{\left(-\mathrm{i} m \phi\right)}\, \sin\phi\, \mathrm{d}\phi .
\end{aligned}
\end{equation}

The azimuthal integral can be reduced once we apply Fourier orthogonality (\(\sin\phi = 1/(2\mathrm{i})(\exp{\left(\mathrm{i}\phi\right)} - \exp{\left(-\mathrm{i}\phi\right)})\)),
\begin{equation}
    \int_0^{2\pi} \exp{\left(-\mathrm{i} m \phi\right)}\, \sin\phi\, \mathrm{d}\phi = \mathrm{i} \pi\left(-\delta_{m,+1} + \delta_{m,-1}\right),
\end{equation}
which shows that only \(m = \pm 1\) contributes.  
The polar integral can be simplified \cite{prop,Ranha_Neves} as  
\begin{equation}
    \int_0^\pi P_n^m(\cos\theta)\, \exp{\left(\mathrm{i} k r \cos\theta\right)}\, \sin^2\theta\, \mathrm{d}\theta
    = 2\,\mathrm{i}^{n+1}\, \frac{(n+m)!}{(n-m)!}\, \frac{j_n(kr)}{kr}.
\end{equation} 

Thus, the coefficients reduce to
\begin{equation}
    G_{n,\pm 1}^{\text{TE}} = -\, \mathrm{i}^{\,n+2}\, \sqrt{(2n+1)\pi} = \mathrm{i}\,G_n ,
\end{equation}
similarly we obtain
\begin{equation}
    G_{n,\pm 1}^{\text{TM}} =  \mp \, \mathrm{i}^{\,n+1}\, \sqrt{(2n+1)\pi} = \pm\, G_n.
\end{equation}
where
\begin{equation}
    G_n = -\,\mathrm{i}^{\,n+1}\, \sqrt{(2n+1)\pi}.
\end{equation}

\begin{figure}[h]
\centering
\fbox{\includegraphics[width=1\linewidth]{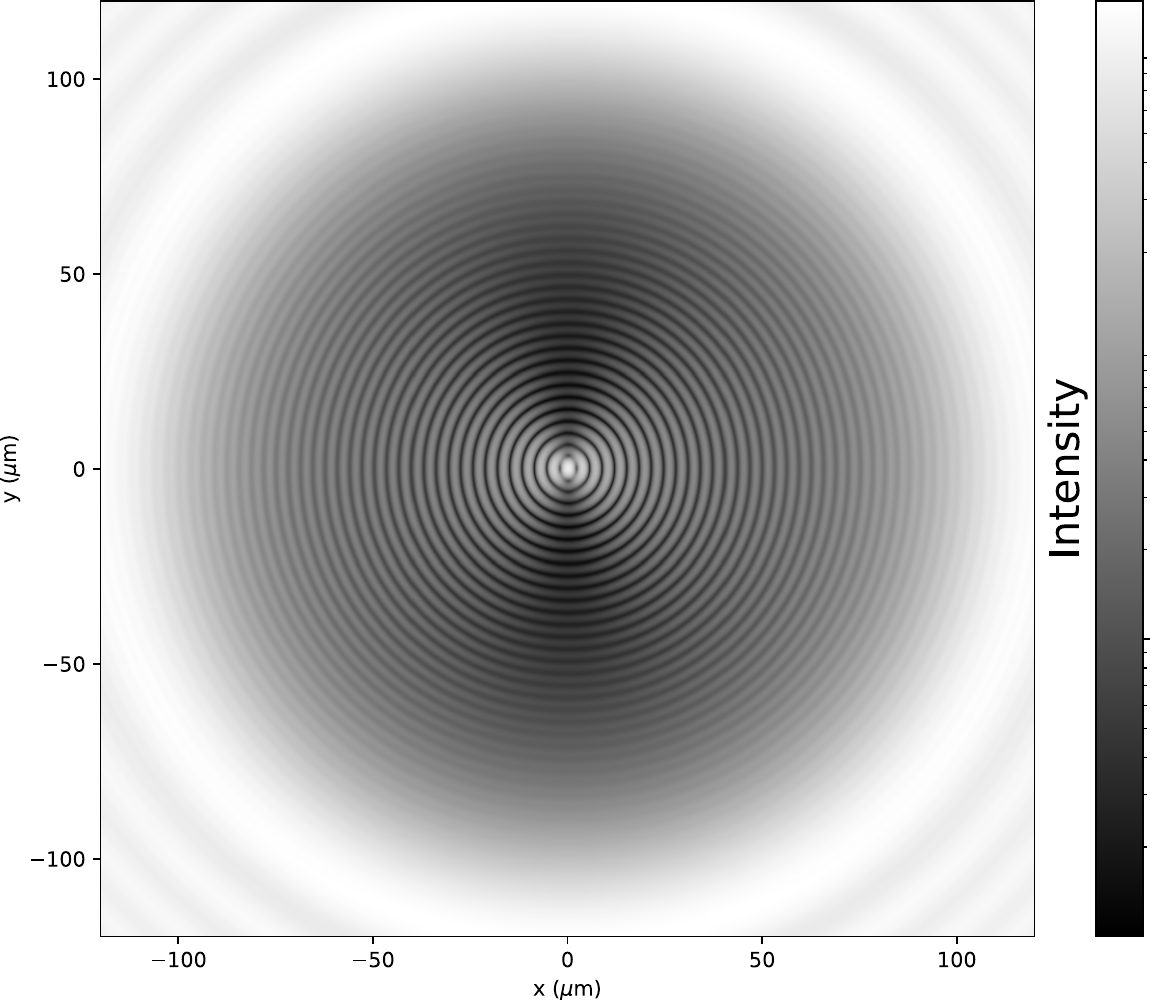}}
\caption{Intensity profile for a sphere with \(R = 100\) \textmu m in Log Scale for intensity. This is the same info from the Fig 2 from the main file, where we have the incident plane wave polarized on the x axis.}
\label{fig:MieLog}
\end{figure}

In this context, we made the image of the profile seen a distance forward. To add better visibility of intensity varying, we put the intensity scale in log, thus giving the Figure \ref{fig:MieLog}. And as an addition, we also made the image of the sintensity in multiple z points forward.

\begin{figure}[h]
\centering
\fbox{\includegraphics[width=1\linewidth]{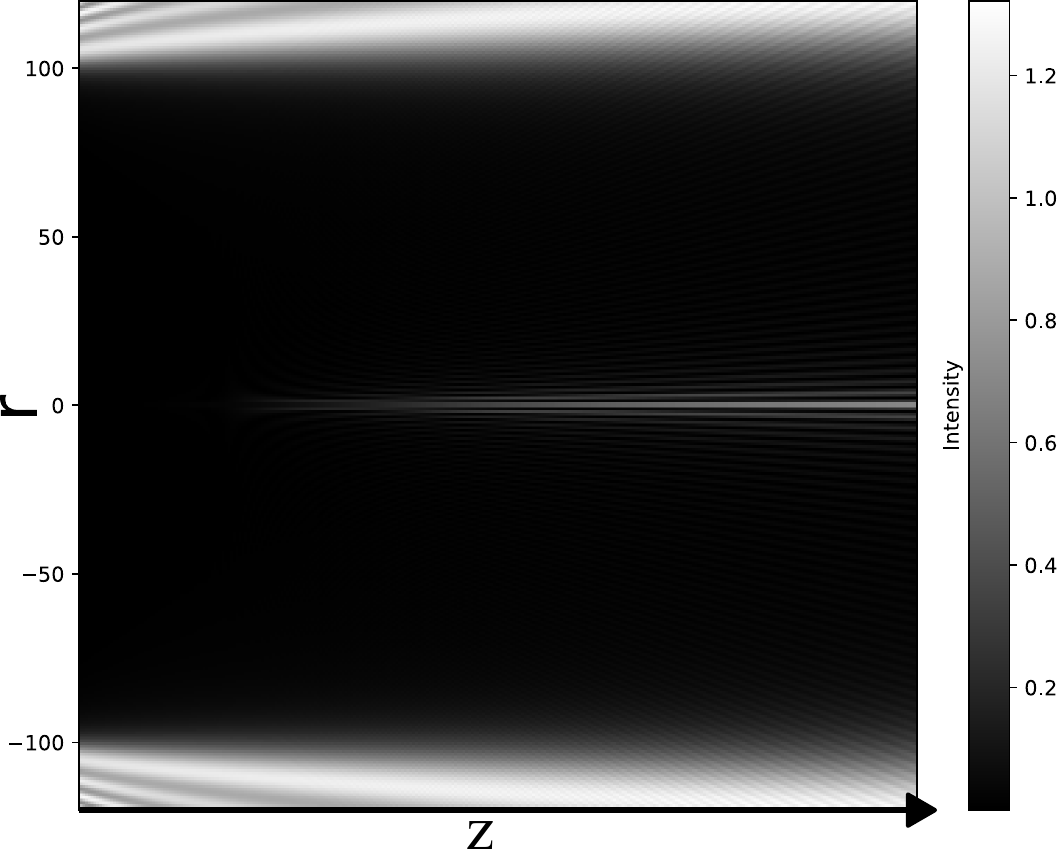}}
\caption{Intensity profile for every radius file giving each z point simulated for a sphere with \(R = 100\) \textmu m and going from R to 1000 R.}
\label{fig:MieZ}
\end{figure}

\newpage
We can also examine the scattering pattern for an incident wave polarized along the y -- axis, for which the Mie coefficients are given by
\begin{equation}
    G_{n,\pm 1}^{\text{TE}, \text{y-pol}} = \pm G_n,
\end{equation}
\begin{equation}
    G_{n,\pm 1}^{\text{TM}, \text{y-pol}} = -\mathrm{i}~ G_n.
\end{equation}

\begin{figure}[h]
\centering
\fbox{\includegraphics[width=0.7\linewidth]{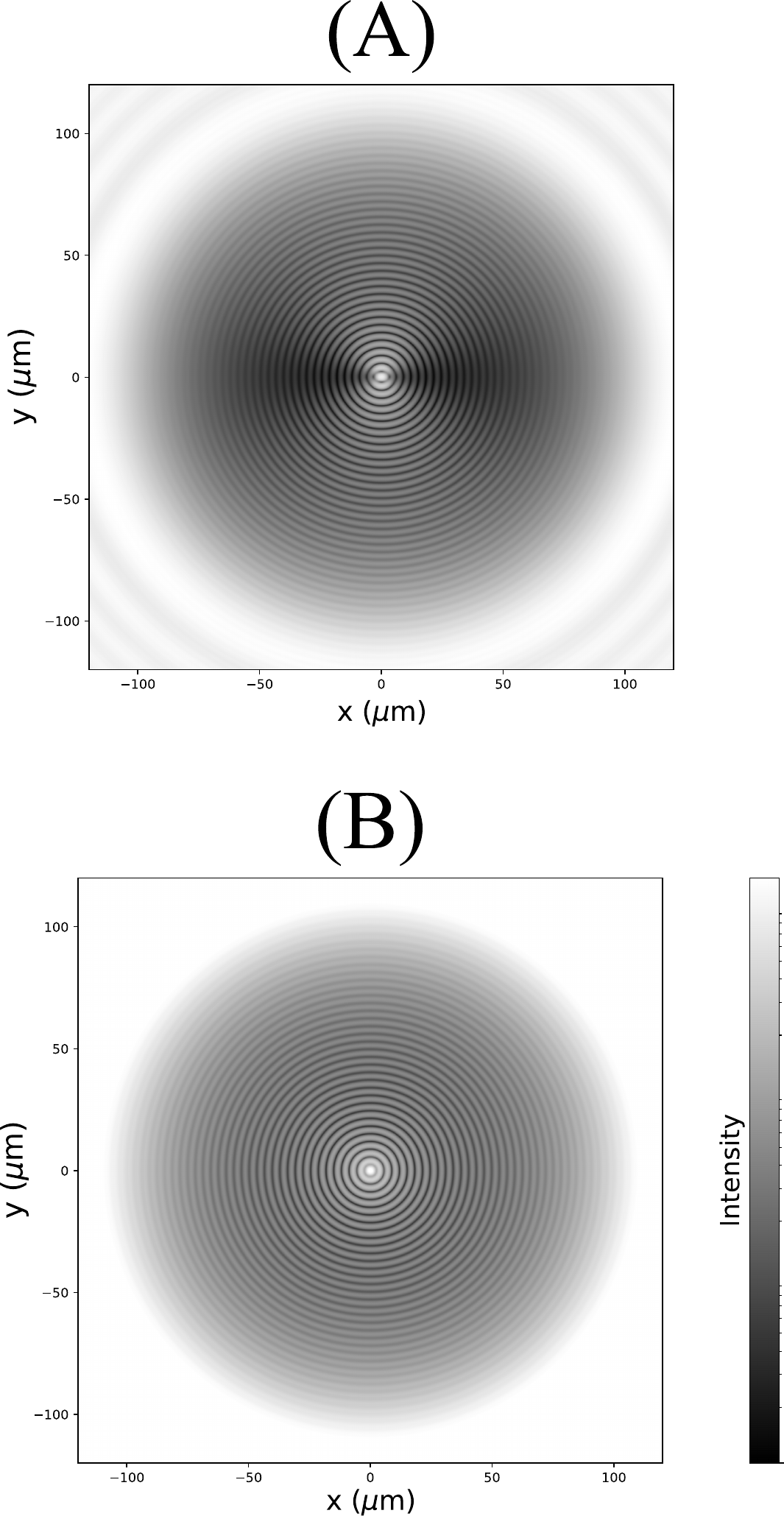}}
\caption{Intensity profile for a sphere with \(R = 100\) \textmu m in Log Scale for intensity at \(z = 1000\) \textmu m. In (A) an y axis and in (B) circular polarized incident plane wave was use.}
\label{fig:Miehaty}
\end{figure}

Using these coefficients, the corresponding intensity pattern was calculated on a logarithmic scale, as shown in Figure~\ref{fig:Miehaty} (A). It can be observed that a slight deformation appears along the x -- axis. To further highlight the vectorial dependence of the scattering, we also computed the pattern for a circularly polarized incident wave, for which the Mie coefficients are

\begin{equation}
    G_{n,\pm 1}^{\text{TE}, \text{circ-pol}} = \mathrm{i}~(1\pm1)~ G_n,
\end{equation}
\begin{equation}
    G_{n,\pm 1}^{\text{TM}, \text{circ-pol}} = (1\pm 1)~G_n.
\end{equation}

In this case, the resulting pattern exhibits no deformation, as shown in Figure~\ref{fig:Miehaty} (B).

\newpage

\subsection{Expanded Electric Fields for plane wave}
We defined the incident electric field as

\begin{equation}
    \textbf{E}_{\text{inc}}= E_0 \sum_{n,m}\left( \frac{\mathrm{i}}{k}G_{n,m}^{\text{TM}}\nabla \times j_n(k r)\textbf{X}_{n,m}(\theta,\phi) + G_{n,m}^{\text{TE}}j_n(k r)\textbf{X}_{n,m}(\theta,\phi) \right),
    \label{E_inc0}
\end{equation}
in the case of a plane wave incident electric field we only have the \(m = \pm1\), substituting this on the field equations gives us
\begin{equation}
    \textbf{E}_{\text{inc}} = E_0\sum_n \frac{\mathrm{i}}{k_1}G_n \left[ \nabla \times j_n(k_1 r) ( \textbf{X}_{n,1}(\theta,\phi) - \textbf{X}_{n,-1}(\theta,\phi))\right] + \mathrm{i}G_n j_n(k_1 r) ( \textbf{X}_{n,1}(\theta,\phi) + \textbf{X}_{n,-1}(\theta,\phi)),
\end{equation}
once we define
\begin{equation}
    \textbf{X}_{n,m}(\theta,\phi) = \frac{1}{\sqrt{n(n+1)}}\textbf{L}Y_{n,m}(\theta,\phi),
\end{equation}
we have
\begin{equation}
    \textbf{X}_{n,1}(\theta,\phi) \pm \textbf{X}_{n,-1}(\theta,\phi) = \frac{1}{\sqrt{n(n+1)}}\textbf{L} (\textbf{Y}_{n,1}(\theta,\phi) \pm \textbf{Y}_{n,-1}(\theta,\phi)),
\end{equation}
being \(\textbf{L}\) the operator
\begin{eqnarray}
    \textbf{L} = -\mathrm{i}\textbf{r}\times\nabla,
\end{eqnarray}
and knowing that 
\begin{equation}
    P_l^{-m} = (-1)^m\frac{(l-m)!}{(l+m)!}P_l^{m},
    \label{eq:Plm}
\end{equation}
we write
\begin{equation}
    Y_{n,-1}(\theta,\phi) = (-1)^mY_{n,1}^*(\theta,\phi),
\end{equation}
leading us to write by applying \(\exp{\left(\mathrm{i}\phi\right)} + \exp{\left(-\mathrm{i}\phi\right)} = 2\cos\phi\) and \(\exp{\left(\mathrm{i}\phi\right)} - \exp{\left(-\mathrm{i}\phi\right)} = 2\mathrm{i}\sin\phi\),
\begin{eqnarray}
    \textbf{X}_{n,1}(\theta,\phi) + \textbf{X}_{n,-1}(\theta,\phi) &=& \textbf{X}_{n,1}(\theta,\phi) - \textbf{X}_{n,1}^*(\theta,\phi) =  \frac{2\mathrm{i}}{n(n+1)}\sqrt{\frac{2n+1}{4\pi}}\textbf{L}(P_n^1(\cos\theta)\sin\phi)\\
    \textbf{X}_{n,1}(\theta,\phi) - \textbf{X}_{n,-1}(\theta,\phi) &=&\textbf{X}_{n,1}(\theta,\phi) + \textbf{X}_{n,1}^*(\theta,\phi) = \frac{2}{n(n+1)}\sqrt{\frac{2n+1}{4\pi}}\textbf{L}(P_n^1(\cos\theta)\cos\phi),
\end{eqnarray}
by applying the definition of \(\textbf{L}\) and already applying the \(\nabla\)
\begin{eqnarray*}
    \textbf{X}_{n,1}(\theta,\phi) + \textbf{X}_{n,-1}(\theta,\phi) &=& \frac{2\mathrm{i}}{n(n+1)}\sqrt{\frac{2n+1}{4\pi}}(-\mathrm{i}) \textbf{r}\times \left[ 
    \hat{\theta}\sin\phi \frac{\partial P_n^1(\cos\theta)}{\partial \theta} + \hat{\phi} \frac{P_n^1(\cos\theta)}{\sin\theta}\cos\phi \right]\\
    &=&+\frac{1}{n(n+1)}\sqrt{\frac{2n+1}{\pi}} \left[ 
    \hat{\phi}\sin\phi \frac{\partial P_n^1(\cos\theta)}{\partial \theta} -\hat{\theta} \frac{P_n^1(\cos\theta)}{\sin\theta}\cos\phi \right],
\end{eqnarray*}
\begin{eqnarray*}
    \textbf{X}_{n,1}(\theta,\phi) - \textbf{X}_{n,-1}(\theta,\phi) &=& \frac{2}{n(n+1)}\sqrt{\frac{2n+1}{4\pi}}(-\mathrm{i}) \hat{r}\times \left[ 
    \hat{\theta}\cos\phi \frac{\partial P_n^1(\cos\theta)}{\partial \theta} - \hat{\phi} \frac{P_n^1(\cos\theta)}{\sin\theta}\sin\phi \right]\\
    &=&-\frac{\mathrm{i}}{n(n+1)}\sqrt{\frac{2n+1}{\pi}} \left[ 
    \hat{\phi}\cos\phi \frac{\partial P_n^1(\cos\theta)}{\partial \theta} + \hat{\theta} \frac{P_n^1(\cos\theta)}{\sin\theta}\sin\phi \right],
\end{eqnarray*}
leading us to write
\begin{eqnarray*}
    \textbf{E}_{\text{inc}} &=& E_0\sum_n \frac{1}{k}\frac{G_n}{n(n+1)}\sqrt{\frac{2n+1}{\pi}} \left[ \nabla \times j_n(k r) \left ( 
    \hat{\phi}\cos\phi \frac{\partial P_n^1(\cos\theta)}{\partial \theta} + \hat{\theta} \frac{P_n^1(\cos\theta)}{\sin\theta}\sin\phi \right)\right] + \\ 
    &+& \frac{\mathrm{i}}{n(n+1)}\sqrt{\frac{2n+1}{\pi}}G_n j_n(k r)  \left[ 
    \hat{\phi}\sin\phi \frac{\partial P_n^1(\cos\theta)}{\partial \theta} -\hat{\theta} \frac{P_n^1(\cos\theta)}{\sin\theta}\cos\phi \right],
\end{eqnarray*}
where remembering that \(\nabla \times(a\textbf{B})=(\nabla a)\times\textbf{B} + a(\nabla\times\textbf{B})\),the first term can be expanded
\begin{eqnarray*}
    \nabla j_n(k r)\times \left( \hat{\phi}\cos\phi \frac{\partial P_n^1(\cos\theta)}{\partial \theta} + \hat{\theta} \frac{P_n^1(\cos\theta)}{\sin\theta}\sin\phi \right) + j_n(k r) \nabla\times \left( \hat{\phi}\cos\phi \frac{\partial P_n^1(\cos\theta}{\partial \theta} + \hat{\theta} \frac{P_n^1(\cos\theta)}{\sin\theta}\sin\phi \right),
\end{eqnarray*}
using
\begin{equation}
    \frac{\mathrm{d} P_n^1(\cos\theta)}{\mathrm{d} \theta} = \frac{1}{\sin\theta} \left( nP_{n+1}^1(\cos\theta) - (n+1)\cos\theta P_n^1(\cos\theta)\right),
    \label{eq:dPn}
\end{equation}
we can write 
\begin{eqnarray*}
    & &\frac{\mathrm{d} j_n(k r)}{\mathrm{d} r}\left[ -\hat{\theta}\frac{\mathrm{d} P_n^1(\cos\theta)}{\mathrm{d} \theta}\cos\phi + \hat{\phi} \frac{P_n^1(\cos\theta)}{\sin\theta}\sin\phi\right] + \\
    &+&\frac{j_n(k r)}{r} [ \frac{1}{\sin\theta}\left(\frac{\mathrm{d}}{\mathrm{d} \theta}\left( nP_{n+1}^1(\cos\theta) - (n+1)\cos\theta P_n^1(\cos\theta)\right) - \frac{P_n^1(\cos\theta)}{\sin\theta} \right)\cos\phi\hat{r} +\\
    &-& \cos\phi \frac{\mathrm{d} P_n^1(\cos\theta)}{\mathrm{d} \theta}\hat{\theta} + \frac{\sin\phi}{\sin\theta}P_n^1(\cos\theta) \hat{\phi} ],
\end{eqnarray*}
by applying the property
\begin{eqnarray}
    \frac{\mathrm{d}}{\mathrm{d} \theta}P_{n+1}^1(\cos\theta)= \frac{1}{\sin\theta}\left((n+1)\cos\theta P_{n+1}^1(\cos\theta) -(n+2)P_{n}^1(\cos\theta)\right)
\end{eqnarray}
we have in the equation above
\begin{eqnarray*}
     & &\frac{\mathrm{d} j_n(k r)}{\mathrm{d} r}\left[ -\hat{\theta}\frac{\mathrm{d} P_n^1(\cos\theta)}{\mathrm{d} \theta}\cos\phi + \hat{\phi} \frac{P_n^1(\cos\theta)}{\sin\theta}\sin\phi\right] 
    +\\
    &-&\frac{ j_n(k r)}{ r}\cos\phi P_n^1(\cos\theta)n(n+1) +\frac{j_n(k r)}{r} \left[  -\cos\phi \frac{\partial P_n^1(\cos\theta)}{\partial \theta}\hat{\theta} + \frac{\sin\phi}{\sin\theta}P_n^1(\cos\theta) \hat{\phi} \right],
\end{eqnarray*}
Allowing us to write the incident field
\begin{eqnarray*}
    \textbf{E}_{\text{inc}} &=& E_0\sum_n \frac{G_n}{n(n+1)}\sqrt{\frac{2n+1}{\pi}}\{  \frac{1}{r k} [  -n(n+1) j_n(k r)\cos\phi P_n^1(\cos\theta)\hat{r} +\\ &-&\frac{\mathrm{d} }{\mathrm{d} r}(r j_n(k r))\frac{\mathrm{d} P_n^1(\cos\theta)}{\mathrm{d} \theta}\cos\phi\hat{\theta} +   \frac{\mathrm{d} }{\mathrm{d} r}(r j_n(k r))\frac{P_n^1(\cos\theta)}{\sin\theta}\sin\phi\hat{\phi}\ ] +\\
    &+&\mathrm{i} j_n(k r)  \left[ 
    \hat{\phi}\sin\phi \frac{\mathrm{d} P_n^1(\cos\theta)}{\mathrm{d} \theta} -\hat{\theta} \frac{P_n^1(\cos\theta)}{\sin\theta}\cos\phi \right] \},
\end{eqnarray*}
that can be rearranged as
\begin{eqnarray*}
    \small
    \textbf{E}_{\text{inc}} &=& E_0\sum_n \frac{G_n}{n(n+1)}\sqrt{\frac{2n+1}{\pi}}\left[ -\cos\phi\left(\frac{1}{k r} j_n^{(1)}(kr)P_n^1(\cos\theta) n(n+1)   \right)\hat{r} \right. \\
     &-&\left. \cos\phi  \left( \frac{1}{ k r} \frac{\mathrm{d}}{\mathrm{d} r}(r j_n^{(1)}(k r))\frac{\mathrm{d} P_n^1(\cos\theta)}{\mathrm{d}\theta} + \mathrm{i}j_n^{(1)}(k r)\frac{P_n^1(\cos\theta)}{\sin\theta} \right)\hat{\theta} \right. \\
    &+&\left.\sin\phi \left(\frac{1}{ k r} \frac{\mathrm{d}}{\mathrm{d} r}(r j_n^{(1)}(k r))\frac{P_n^1(\cos\theta)}{\sin\theta} +\mathrm{i}j_n^{(1)}(k r)\frac{\mathrm{d} P_n^1(\cos\theta)}{\mathrm{d}\theta} \right)\hat{\phi}\right].
\end{eqnarray*}

Analogously, we can obtain the scattered field.

\begin{eqnarray*}
    \small
    \textbf{E}_{\text{sca}} &=& E_0\sum_n \frac{G_n}{n(n+1)}\sqrt{\frac{2n+1}{\pi}}\left[ -\cos\phi\left(\frac{a_n}{k r} h_n^{(1)}(kr)P_n^1(\cos\theta) n(n+1)   \right)\hat{r} \right. \\
     &-&\left. \cos\phi  \left( \frac{a_n}{ k r} \frac{\mathrm{d}}{\mathrm{d} r}(r h_n^{(1)}(k r))\frac{d P_n^1(\cos\theta)}{d\theta} + \mathrm{i}b_nh_n^{(1)}(k r)\frac{P_n^1(\cos\theta)}{\sin\theta} \right)\hat{\theta} \right. \\
    &+&\left.\sin\phi \left(\frac{a_n}{ k r} \frac{\mathrm{d}}{\mathrm{d} r}(r h_n^{(1)}(k r))\frac{P_n^1(\cos\theta)}{\sin\theta} +\mathrm{i}b_nh_n^{(1)}(k r)\frac{\mathrm{d} P_n^1(\cos\theta)}{\mathrm{d}\theta} \right)\hat{\phi}\right],
\end{eqnarray*}
where $a_n = {a_{n,1}}/G_{n}$ and $b_n = {b_{n,-1}}/G_{n}$.

\subsection{Intensity at the center for plane wave}

Still in the situation of a plane wave, we can write the scattered electric field as
\begin{eqnarray*}
    \small
    \textbf{E}_{\text{sca}} &=& -E_0\sum_n \frac{\mathrm{i}^{n+1}(2n+1)}{n(n+1)}\left[ -\cos\phi\left(\frac{a_n}{k r} h_n^{(1)}(kr)P_n^1(\cos\theta) n(n+1)   \right)\hat{r} \right. \\
     &-&\left. \cos\phi  \left( \frac{a_n}{ k r} \frac{\mathrm{d}}{\mathrm{d} r}(r h_n^{(1)}(k r))\frac{d P_n^1(\cos\theta)}{d\theta} + \mathrm{i}b_nh_n^{(1)}(k r)\frac{P_n^1(\cos\theta)}{\sin\theta} \right)\hat{\theta} \right. \\
    &+&\left.\sin\phi \left(\frac{a_n}{ k r} \frac{\mathrm{d}}{\mathrm{d} r}(r h_n^{(1)}(k r))\frac{P_n^1(\cos\theta)}{\sin\theta} +\mathrm{i}b_nh_n^{(1)}(k r)\frac{d P_n^1(\cos\theta)}{d\theta} \right)\hat{\phi}\right].
\end{eqnarray*}
Which is a complex equation, but in the case of the intensity at the center of the circle through multiple distances forwards it can be simplified by applying the \(\theta = 0\), allowing the use of the properties
\begin{eqnarray}
    \lim_{\theta\to 0} \frac{P_n^1(\cos\theta)}{\sin\theta}     &=& -\frac{1}{2}n(n+1),\\ \lim_{\theta\to 0} \frac{\mathrm{d} P_n^1(\cos\theta)}{\mathrm{d}\theta}     &=& -\frac{1}{2}n(n+1),\\ 
    \lim_{\theta\to 0}P_n^1(\cos\theta) &=& 0,
\end{eqnarray}
considering that in this case, $r$ becomes $z$, we can rewrite it as,
\begin{eqnarray*}
    \small\textbf{E}_{\text{sca}} &=& E_0\sum_n \frac{\mathrm{i}^{n+1}}{2}(2n+1)\left[ -\cos\phi  \left( \frac{a_n}{ k z} \frac{\mathrm{d}}{\mathrm{d} z}(z h_n^{(1)}(k r)) + \mathrm{i}b_nh_n^{(1)}(k z) \right)\hat{\theta} +\right. \\
    &+&\left.\sin\phi \left(\frac{a_n}{ k z} \frac{\mathrm{d}}{\mathrm{d} z}(z h_n^{(1)}(k z)) +\mathrm{i}b_nh_n^{(1)}(k z) \right)\hat{\phi}\right].
\end{eqnarray*}

From where we can have any value of \(\phi\), so letting \(\phi = 0\) give us
\begin{eqnarray}
    \small\textbf{E}_{\text{sca}} = -E_0\sum_n \frac{i^{n+1}}{2}(2n+1)  \left( \frac{a_n}{ k z} \frac{\mathrm{d}}{\mathrm{d} z}(r h_n^{(1)}(k z)) + \mathrm{i}b_nh_n^{(1)}(k z) \right)\hat{\theta},
\end{eqnarray}
from where we can rewrite the equation by applying the derivative, as 
\begin{equation}
\frac{d }{d r} \left( r h_n^{(1)}(kr) \right) = h_n^{(1)}(kr) + r\frac{d}{d r}\left( h_n^{(1)}(kr)\right),
\end{equation}
together with the fact that 
\begin{equation}
     \frac{d}{dr}h_n^{(1)}(kr) = \frac{n}{r}\cdot h_n^{(1)}(kr) - k\cdot h_n^{(1)}(kr),
\end{equation}
so we end up with
\begin{eqnarray}
    \small\textbf{E}_{\text{sca}} = -E_0\sum_n \frac{\mathrm{i}^{n+1}}{2}(2n+1)  \left[ \frac{a_n}{ k z} \left( (n+1)\cdot h_n^{(1)}(kz) - kz \cdot h_{n+1}^{(1)}(kz) \right) + \mathrm{i}b_nh_n^{(1)}(k z) \right]\hat{\theta},
\end{eqnarray}
and for the incident field
\begin{eqnarray}
    \small\textbf{E}_{\text{inc}} = E_0 \cdot \exp{\left(\mathrm{i} kz\right)}\hat{x} .
\end{eqnarray}

\begin{figure}[h]
\centering
\fbox{\includegraphics[width=1\linewidth]{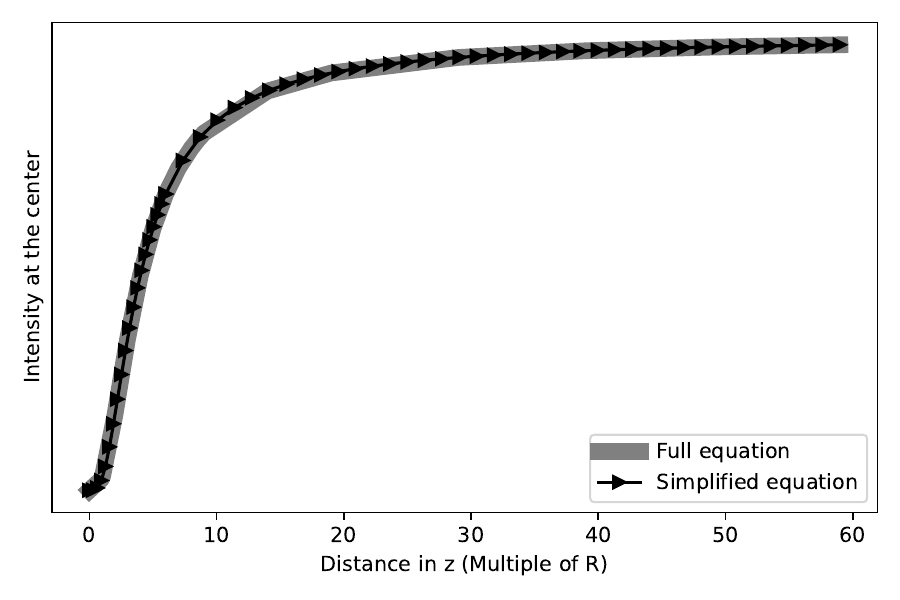}}
\caption{Comparison of the axial intensity (\(x=0, y=0\)) along the \(z\)-axis for a sphere of radius \(R = 100\) \textmu m. The results from the full Mie scattering calculation (solid line) are compared with the simplified expression (dashed line), showing excellent agreement.}
\label{fig:Izs}
\end{figure}

To illustrate the validity of the simplified expression of fields to use in the calculations, we consider the case of a sphere with radius \(R = 100\) \textmu m. Figure~\ref{fig:Izs} compares the intensity along the optical axis (\(x=0, y=0\)) obtained from the full Mie scattering calculation with that obtained from the simplified formulation. The results show excellent agreement, confirming that the simplified approach reproduces the same axial intensity profile as the full calculation.

Using these simplified fields, we can describe the total electric field, whose squared modulus is
\begin{equation}
    |\textbf{E}_t|^2 = (\textbf{E}_{\text{inc}}+\textbf{E}_{\text{scat}})(\textbf{E}^*_{\text{inc}}+\textbf{E}^*_{\text{scat}}),
\end{equation}
which can be expanded as

\begin{eqnarray}
    |\textbf{E}_t|^2 = \textbf{E}_{\text{inc}}\cdot\textbf{E}^*_{\text{inc}}  + \textbf{E}_{\text{sca}}\cdot\textbf{E}^*_{\text{sca}}  + \textbf{E}_{\text{inc}}\cdot\textbf{E}^*_{\text{sca}}  + \textbf{E}^*_{\text{inc}}\cdot\textbf{E}_{\text{sca}} ,
\end{eqnarray}
and giving the fact that \(\textbf{b}\cdot\textbf{a}^*+\textbf{a}\cdot\textbf{b}^* = 2\text{Re}\left(\textbf{a}\cdot\textbf{b}^*\right)\), we can write
\begin{eqnarray}
    |\textbf{E}_t|^2 = \textbf{E}_{\text{inc}}\cdot\textbf{E}^*_{\text{inc}}  + \textbf{E}_{\text{sca}}\cdot\textbf{E}^*_{\text{sca}}  + 2\text{Re}\left(\textbf{E}_{\text{sca}}\cdot\textbf{E}^*_{\text{inc}} \right).
\end{eqnarray}

\begin{figure}[h]
\centering
\fbox{\includegraphics[width=1\linewidth]{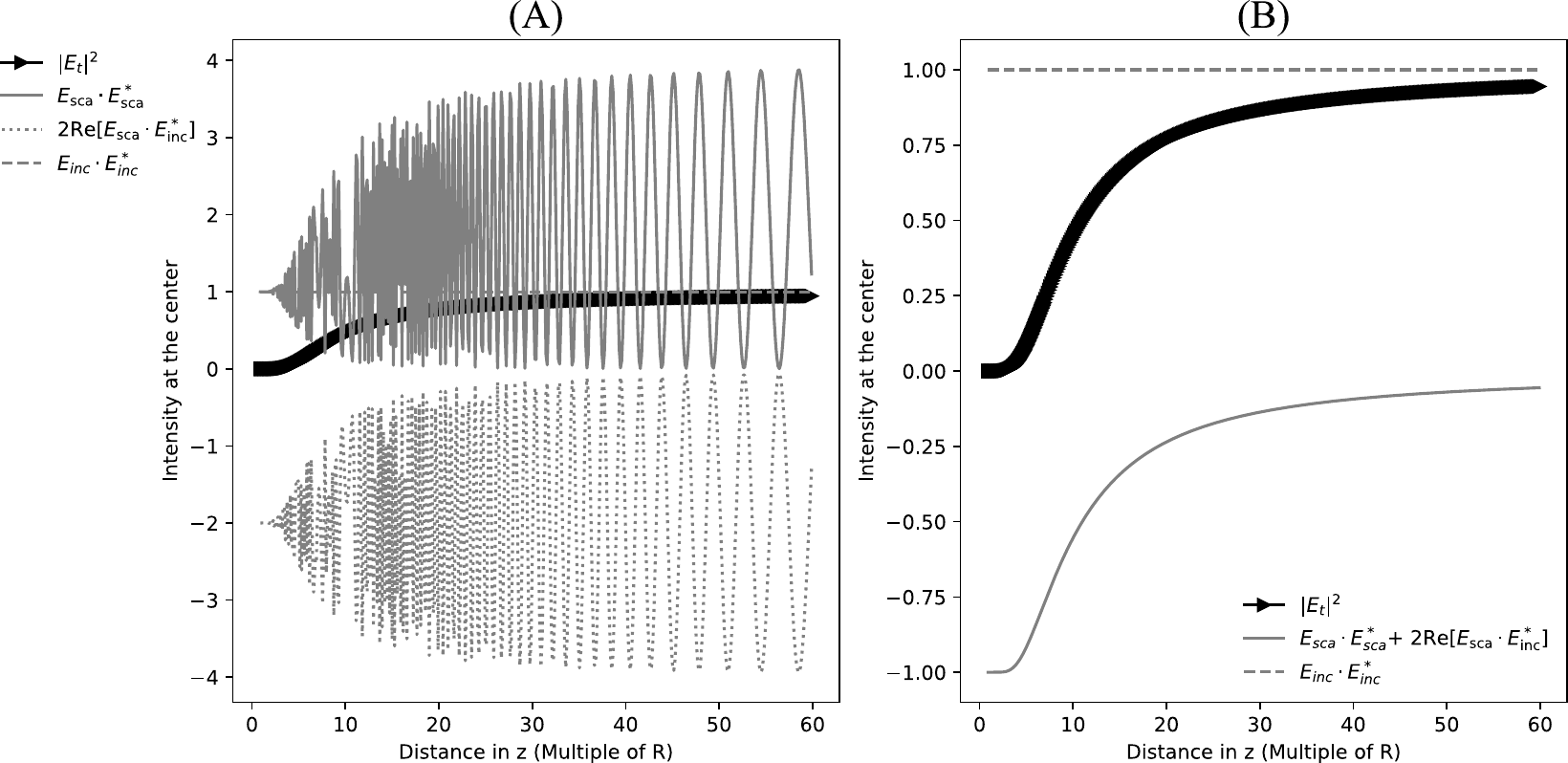}}
\caption{Comparison of each term of the total field modulus. (A) Each term separately, the incident field, the scattered and the sum of the product between the incident and scattered. (B) The incident field term versus the scattered and cross terms. This was made for the case of a \(R = 1000\) \textmu m.}
\label{fig:Terms}
\end{figure}

And writing this cross term we have
\begin{equation}
    \textbf{E}_{\text{sca}}\cdot\textbf{E}^*_{\text{inc}} = -\exp{\left(-\mathrm{i} k z\right)} \cdot  \sum_n \frac{\mathrm{i}^{n+1}}{2}(2n+1)  \left[ \frac{a_n}{ k z} \left( (n+1)\cdot h_n^{(1)}(kz) - kz \cdot h_{n+1}^{(1)}(kz) \right) + \mathrm{i}b_nh_n^{(1)}(k z) \right],
\end{equation}
This term is challenging to simplify analytically. However, by examining their contributions to the total intensity, we can make several observations. First, the modulus of the scattered field is shifted relative to the sum of these cross-product terms, as shown in Figure \ref{fig:Terms}(A). Second, if we sum the three terms corresponding to the scattered and cross terms, we recover a profile that matches the total intensity curve, but shifted to negative values because it combines with the incident field modulus, which is normalized to 1. This behavior is illustrated in Figure \ref{fig:Terms}(B).

\newpage

\subsection{Method to Differentiate Circle from Sphere}

\begin{figure}[h]
\centering
\includegraphics[width=1\linewidth]{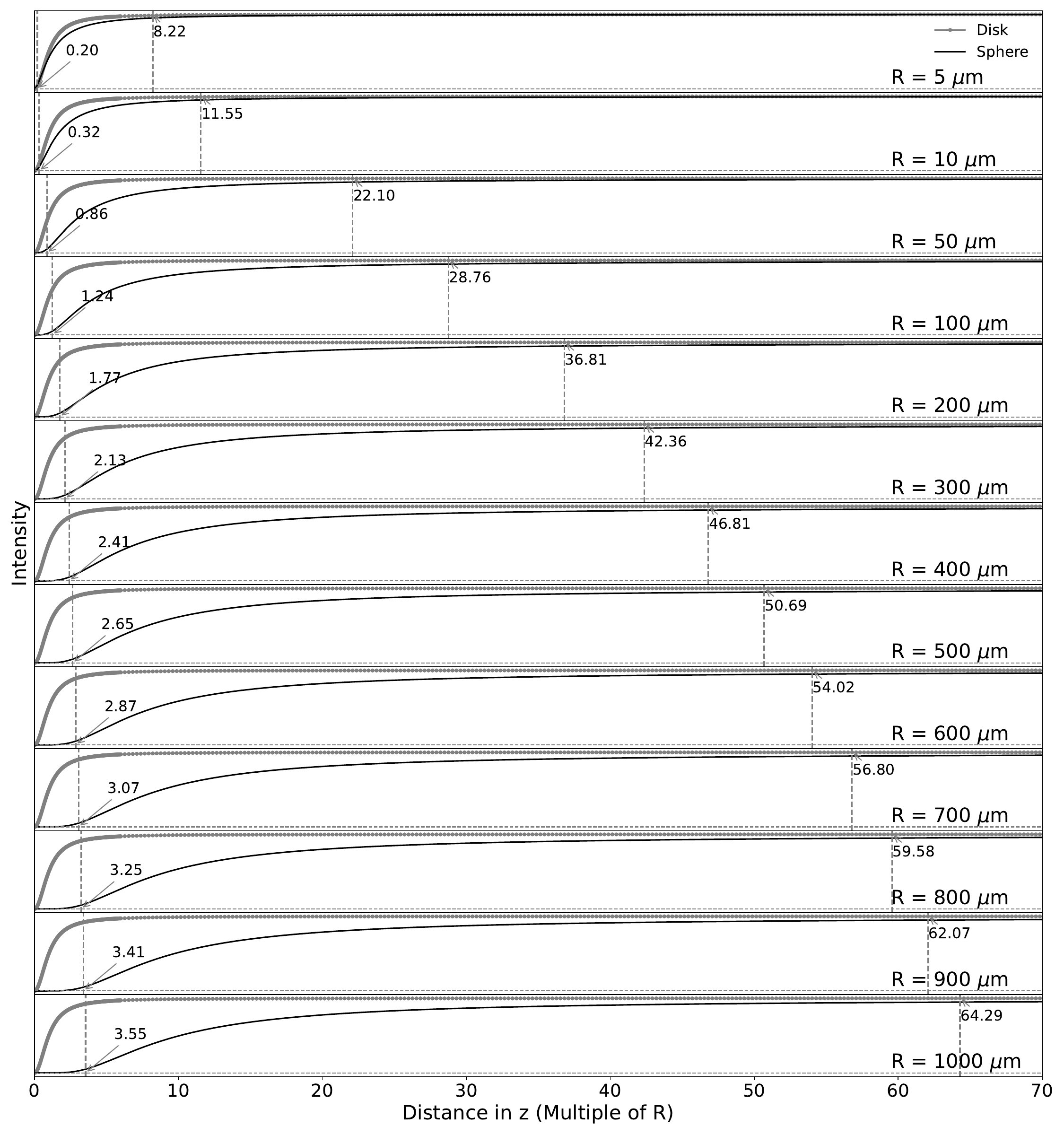}
\caption{Intensity along the z--axis, for different scatterer sizes, both disc and sphere with same radius. Two points are evidence, in dotted lines, for 5\% and 95\% of full intensity of the sphere intensity profile.}
\label{fig:Izrs}
\end{figure}

To better understand the difference between a sphere and a disk, the central intensity was plotted as a function of z for several radius, as shown in Figure \ref{fig:Izrs}. It can be seen that for very small radii, the intensity patterns of the sphere and the disk are similar, while increasing the radius leads to more pronounced differences. The points where the intensity first rises from zero and where it reaches the plateau are also marked. By examining these points normalized to the same distance ratio for each sphere, no direct correlation with the radius was observed; however, a general trend of increasing separation with larger radius is evident. And specially looking at the intensity at which it reaches 95\% of the full intensity we can clearly differentiate them as the disk generally emerges for $z$ distances smaller them the radius of disk.

\FloatBarrier

\bibliography{sample}

\end{document}